\documentclass[aps,prl,twocolumn,amsmath,amssymb]{revtex4}
\usepackage{graphicx,epsfig,epsf}
\usepackage{dcolumn}
\usepackage{bm}

\begin{document}
\newcommand{\ri}{{\rm i}}
\newcommand{\re}{{\rm e}}
\newcommand{\bx}{{\bf x}}
\newcommand{\bd}{{\bf d}}
\newcommand{\br}{{\bf r}}
\newcommand{\bk}{{\bf k}}
\newcommand{\bE}{{\bf E}}
\newcommand{\bR}{{\bf R}}
\newcommand{\bM}{{\bf M}}
\newcommand{\bn}{{\bf n}}
\newcommand{\bs}{{\bf s}}
\newcommand{\tbs}{\tilde{\bf s}}
\newcommand{\rSi}{{\rm Si}}
\newcommand{\beps}{\mbox{\boldmath{$\epsilon$}}}
\newcommand{\rg}{{\rm g}}
\newcommand{\tr}{{\rm tr}}
\newcommand{\xmax}{x_{\rm max}}
\newcommand{\ra}{{\rm a}}
\newcommand{\rx}{{\rm x}}
\newcommand{\rs}{{\rm s}}
\newcommand{\rP}{{\rm P}}
\newcommand{\up}{\uparrow}
\newcommand{\down}{\downarrow}
\newcommand{\hc}{H_{\rm cond}}
\newcommand{\kb}{k_{\rm B}}
\newcommand{\cI}{{\cal I}}
\newcommand{\tit}{\tilde{t}}
\newcommand{\cE}{{\cal E}}
\newcommand{\cC}{{\cal C}}
\newcommand{\Ubs}{U_{\rm BS}}
\sloppy

\title{Decoherence in a system of many two--level atoms} 
\author{Daniel Braun}
\affiliation{Laboratoire de Physique Th\'eorique, IRSAMC, UMR 5152 du CNRS,
  Universit\'e Paul Sabatier, 118, route de  
  Narbonne, 31062 Toulouse, FRANCE} 

\begin{abstract}
I show that the decoherence in a system of $N$ degenerate two--level atoms
  interacting with a bosonic heat bath is for any number of atoms $N$
  governed 
  by a generalized Hamming distance (called ``decoherence metric'')
  between the superposed quantum states, with a time--dependent metric
  tensor that is specific for the heat bath.   
The decoherence  metric allows
  for the complete characterization of the 
  decoherence of all possible superpositions of 
  many-particle states, and can be applied to 
  minimize the over-all decoherence in a quantum memory. For qubits
  which are far apart, 
  the decoherence is given by a function describing single-qubit decoherence
  times the standard Hamming distance. I apply the theory to cold atoms in
  an optical lattice interacting with black body radiation.
\end{abstract}
\maketitle
Decoherence is considered one of the main ingredients in the transition from
quantum mechanics to classical mechanics \cite{Giulini96}. According to
experimental evidence (see for example 
\cite{Brune96,Vion02})
``decoherence'' can be well described within the standard framework of
quantum mechanics \cite{Weiss93}. The
loss of coherence of quantum mechanical superpositions becomes increasingly
  rapid when the 
``distance'' between the components of the superpositions gets
large \cite{Zurek81,Strunz02}, such that quantum mechanical behavior
like interference of wave functions is never observed for macroscopic
objects. With few notable 
exceptions \cite{Duer02,KrojanskiS04,Mintert05}, most work on decoherence has 
focused so far on systems with only one or a few effective degrees of freedom
coupled to a heat bath. Even when large numbers of 
particles come into play, like in superconducting qubits based on Josephson
  junctions  
\cite{Makhlin01} or superradiance \cite{Bonifacio71a,PBraun98a},
only a single effective degree of freedom is normally retained.
With the rise of quantum information science the fundamental question
whether for large particle (qubit) numbers there might be new
self-limitations of 
quantum mechanics \cite{Martini05,Auffeves03,Lamine06}, based e.g.~on the
amount of 
entanglement in the quantum 
states \cite{Bandyopadhyay05,Cai05}, has also become of huge
practical importance. Indeed, understanding and possibly suppressing
decoherence of quantum superpositions in the exponentially large Hilbert
space of possibly thousands of qubits is considered the most
difficult problem on the way to a large scale quantum computer.

In the following I show that a generalized Hamming distance with a
time--dependent metric tensor given by the heat bath and the couplings to
the heat 
bath completely determines the decoherence process of a
an arbitrarily large number of qubits. The
standard Hamming distance, which counts the number of bits by which two
code words differ, plays a 
central role in  both classical and quantum error correction
\cite{Shor95,Steane96}, and determines in particular what code
words remain distinguishable in a given error model. It is therefore very
satisfying that the heat bath itself determines the metric relevant
for the decoherence process, as decoherence reflects to what extent the heat
bath distinguishes the superposed states \cite{Zurek81}. 
 
Starting point of the analysis are 
$N$ two--level atoms at
arbitrary but fixed positions $\bR_i$ ($i=0,\ldots, N-1$) interacting with
a common bosonic heat bath. All atoms are assumed identical with level spacing
$\hbar \Omega_0$ and 
energy eigenstates $|-1\rangle$ and $|1\rangle$, with $\sigma_z|\pm
1\rangle=\pm |\pm 1\rangle$. 
The total Hamiltonian reads 
\begin{eqnarray}
H&=&\sum_k \hbar\omega_k a_k^\dagger
a_k+\frac{1}{2}\hbar\Omega_0\sum_{i=0}^{N-1}\sigma_{zi}\label{H}\\
&&+\hbar
\sum_k\sum_{i=0}^{N-1}g_k^{(i)}\sigma_{xi}\left(a_k\re^{\ri
  \bk\cdot\bR_i}+a_k^\dagger\re^{-\ri\bk\cdot\bR_i}\right)\,, \nonumber
\end{eqnarray}
where $\sigma_{xi}$ and $\sigma_{zi}$ are Pauli matrices for atom $i$.
While the results obtained are valid for a general bosonic heat bath, we
will consider in the following as concrete example thermal black body
(BBR) radiation dipole-coupled to the atoms \cite{Scully97}.
 The index $k$ then
stands for wave  
vector $\bk$ and polarization 
direction $\lambda$ ($k_j=2\pi n_j/L $ with 
integer $n_j$, $j=x,y,z$ for periodic boundary conditions) of the electro
magnetic waves;
 $a_k^\dagger$ ($a_k$) are the creation 
(annihilation) operators for mode $k$  with frequency $\omega_k=c|\bk|$,
 polarization vector $\epsilon_k$, and electric field amplitude
 $\cE_k=\sqrt{\hbar\omega_k/(2\varepsilon_0 V)}$, where $\varepsilon_0$, $c$,
 and $V$ are the
 dielectric constant
of the vacuum, speed of light, and the quantization volume, respectively.
 The coupling constant of 
 atom $i$ to mode $k$ is denoted by
 $g_k^{(i)}=-\frac{ed\cE_k}{\hbar}\hat{u}^{(i)}\cdot\epsilon_k$, where 
 $\hat{u}^{(i)}$ stands for a unit vector 
in the direction of the dipole moment of atom $i$, $\langle -1|\bd|1\rangle=ed
\hat{u}^{(i)}$ with electron charge $e$ and dipole length $d$.
Hamiltonians of the form (\ref{H}) have been studied before in many
situations, in particular in the context of dipole--dipole interactions
in quantum optics and various forms of mode selection (see \cite{Goldstein96}
and references therein). 
We will restrict ourselves
to atoms with degenerate 
energy levels, $\Omega_0=0$. This should set a lower limit for decoherence
in quantum
memories based on qubits with an allowed dipole transition. From a
theoretical perspective 
degenerate levels are extremely 
attractive, as the model can be solved exactly 
for any $N$.  The results remain valid for small but finite $\Omega_0$, as
long as the time $t\ll 1/(N\hbar\Omega_0)$
\cite{Strunz02}. The 
results are also easily extended to any model with commuting system and
interaction Hamiltonian, $[H_{\rm sys},H_{\rm int}]=0$, such as pure
dephasing models with $H_{\rm sys}\propto\sigma_{zi}$ and $H_{\rm
  int}\propto\sigma_{zi}$. 

It is interesting to consider the decoherence process of $n$ selected
atoms ($i=0,\ldots,n-1$) out of the $N$ atoms. This offers the additional
information how decoherence 
scales as function of the size of a sub-cluster in the quantum
memory. The common heat bath can lead to entanglement between the
atoms \cite{Braun02,Braun05}, such that unobserved atoms can become an
important source of ``indirect'' decoherence. 
Besides its conceptual importance, this can be relevant 
practically, if additional atoms are trapped in an optical lattice, or when
information is stored in an atomic gas in a large number of atoms
\cite{Juulsgaard04}. We therefore first trace out the
bosonic modes and secondly the additional atoms $n\ldots N-1$. The 
reduced density matrix $\rho$ of the remaining atoms will
be expressed in the eigenbasis of the $\sigma_{xi}$, the natural basis (also
called pointer basis) for
studying the decoherence process \cite{Zurek81}. The matrix elements 
of $\rho$ are $\rho_{\tbs\tbs'}(t)$, where
$\tbs=(s_0,\ldots,s_{n-1})$ is a subset of the code word ${\bf
  s}=(s_0,s_1,\ldots s_{N-1})$ (with
$\sigma_{xi}|s_i\rangle_x=s_i|s_i\rangle_x$, $s_i=\pm 1$, and similarly for
$\tbs'$, i.e.~$\tbs$ and $\tbs'$ are binary representations of the matrix
indices). We assume that the heat  
bath is initially 
in thermal equilibrium at temperature $T$, and all atoms are prepared in
a pure state.

Using the
method of shifted harmonic oscillators \cite{Mahan90} one shows that the
time evolution of $\rho$ is given by  
\begin{eqnarray}
\rho_{\tbs\tbs'}(t)&=&\rho_{\tbs\tbs'}(0)\times\label{rhos0}\\
&&\exp\bigg\{-\sum_{i,j=0}^{n-1}(s_i-s_i')(s_j-s_j')f_{ij}(t,\bR_i-\bR_j)\nonumber\\ 
&&+\ri\sum_{j=1}^{n-1}\sum_{i=0}^{j-1}(s_is_j-s_i's_j')\varphi_{ij}(t,\bR_i-\bR_j)\bigg\}\nonumber\\   
&&\prod_{l=n}^{N-1}\cos\left(
\sum_{i=0}^{n-1}(s_i-s_i')\varphi_{il}(t,\bR_i-\bR_l)\right)\,.\nonumber   
\end{eqnarray}
The functions 
$f_{ij}$ and $\varphi_{ij}$ are given by
\begin{eqnarray}
f_{ij}(t,\bR)&=&\sum_k\frac{g_k^{(i)}g_k^{(j)}}{\omega_k^2}\cos(\bk\cdot\bR)
(1-\cos\omega_kt)\coth\frac{\beta\hbar\omega_k}{2}\nonumber\\
\varphi_{ij}(t,\bR)&=&2\sum_k\frac{g_k^{(i)}g_k^{(j)}}{\omega_k^2}\cos(\bk\cdot\bR)(\omega_kt- 
\sin\omega_kt)\label{phij}\,.  
\end{eqnarray}
They both vanish at $t=0$. 

We are interested in the initial stages of the decay 
of the off-diagonal matrix elements (the ``coherences''),
when $f_{ij}$ and $\varphi_{ij}$ are both much smaller than one, as later basically
no coherences are left anyway. I define the ``decoherences''
$d_{\tbs\tbs'}(t)$ through
\begin{eqnarray}
\frac{|\rho_{\tbs\tbs'}(t)|}{|\rho_{\tbs\tbs'}(0)|}&\equiv&1-
d_{\tbs\tbs'}(t)\,.
\end{eqnarray}
Expanding eq.(\ref{rhos0}) to lowest order in
$f_{ij}$ and $\varphi_{ij}$ we find 
\begin{eqnarray}
d_{\tbs\tbs'}(t)&\simeq&\frac{1}{4}(\tbs-\tbs')
\bM(t)(\tbs-\tbs')^T \label{ct}\,,\label{dd} 
\end{eqnarray}
where $^T$ denotes the transpose, and $\bM(t)$ is a real, symmetric, and
non--negative matrix with elements ($i,j=0,\ldots,n-1$)
\begin{eqnarray}
M_{ij}&=&4f_{ij}(t,\bR_i-\bR_j)+2\Phi_{ij}(t,\bR_i,\bR_j)\,,\label{Mij}\\
\Phi_{ij}(t,\bR_i,\bR_j)&=&\sum_{l=n}^{N-1}
\varphi_{il}(t,\bR_i-\bR_l)\varphi_{jl}(t,\bR_j-\bR_l)\,.\nonumber 
\end{eqnarray}
The properties of $\bM$ allow us to consider $\bM(t)$ as a time--dependent
metric tensor, 
and to define a metric in the vector space 
${\mathbb R}^n$ containing all code words. The heat--bath itself therefore
induces a natural distance 
defined as
\begin{equation} \label{dM}
||\tbs-\tbs'||_{M(t)}=\frac{1}{2}\sqrt{(\tbs-\tbs')\bM(t)(\tbs-\tbs')^T}\,.
\end{equation}
We will call $||\tbs-\tbs'||_{M(t)}$ with $\bM(t)$ given by eq.(\ref{Mij})
``decoherence metric'' and $\bM$ ``decoherence metric tensor'' (DMT).
Eq.(\ref{dd}) shows that the decoherence process of a superposition of code
words $\tbs$ and $\tbs'$ is governed by the decoherence metric,
\begin{equation} \label{dd2}
d_{\tbs\tbs'}(t)\simeq||\tbs-\tbs'||_{M(t)}^2\,,
\end{equation}
with a DMT $\bM(t)$ defined by the heat bath itself and the couplings to the
heat bath. 
Eq.(\ref{dd2}) constitutes the central result of this letter.\\
The distance $||\tbs-\tbs'||_{M(t)}$ generalizes the well--known Hamming
distance 
$D^H(\tbs,\tbs')$, which is defined as the number of 
digits in which $\tbs$ and $\tbs'$ differ. The Hamming distance
$D^H(\tbs,\tbs')$ is recovered from eq.(\ref{dM}) for the trivial metric tensor
$\bM={\bf I}$, where ${\bf I}$ is the identity matrix in $n$ dimensions,
as $D^H(\tbs,\tbs')=||\tbs-\tbs'||_I^2$. As an immediate consequence we find
that the entanglement of a quantum state does not (or at least not fully)
determine the decoherence 
between its components. Indeed, all states related by parallel qubit
flips in the components (such as e.g.~$(|000\rangle+|111\rangle)/\sqrt{3}$ and
$(|100\rangle+|011\rangle)/\sqrt{3}$ for $N=3$) have the    
same entanglement and the same Hamming distance, but the off--diagonal
elements of $\bM$ lead to different time dependent decoherences for these
states (see also FIG.\ref{fig.dofd}).\\
The decoherence metric is in the strict sense a pseudometric. It is
symmetric, non--negative, and obeys the triangle 
inequality, but there can be code words $\tbs$ and $\tbs'$ with
$\tbs\ne\tbs'$ such  
that $||\tbs-\tbs'||_{M(t)}=0$. Such code words constitute a decoherence free
subspace (DFS), which plays an important role in quantum information processing
\cite{Zanardi97,Duan98,Lidar98,Braun98}. For the ease of 
language, I will continue to use the word ``metric'' instead of
``pseudometric'', 
however. The desire to keep the triangle inequality motivates the choice of
the square root in eq.(\ref{dM}), whereas the Hamming distance is
defined without it. The latter still obeys the triangle inequality
 for binary code words, but the triangle inequality can be broken if
 the metric tensor has non--diagonal matrix elements and the metric is 
defined without the square root. The decoherence metric has the form of a
Mahalanobis distance, frequently used in statistical analysis
\cite{Mahalanobis36}.\\ 
The first part of the DMT, $4f_{ij}(t,\bR_i-\bR_j)$ in
eq.(\ref{Mij}), depends on the $n$ 
selected atoms only. It describes the direct decoherence of the atoms due to
their exposure to the heat bath.
The second part, $2\Phi_{ij}(t,\bR_i,\bR_j)$, however, depends on all
atoms, even the non--selected ones, such that this contribution
is not translationally invariant.
$\Phi_{ij}$ describes
the ``indirect decoherence'' which arises 
due to the effective interaction and subsequent correlations or even
entanglement that 
the heat bath induces between the atoms \cite{Braun02,Braun05}.
Remarkably, the total decoherence $d_{\rm tot}(t)$, as defined by
$\sum_{\bs\ne\bs'}|\rho_{\bs\bs'}(t)|/\sum_{\bs\ne\bs'}|\rho_{\bs\bs'}(0)|\equiv 
1-d_{\rm 
  tot}(t)$, involves only the trace of the DMT, $
d_{\rm tot}(t)=\frac{1}{2-2^{-n+1}}\tr
\bM(t)$. In
situations where $\Phi_{ii}$ is absent (e.g.~for $n=N$), the 
total decoherence scales for large $n$ proportional to $n$, as $f_{ii}$ is
independent of $n$; otherwise, the
scaling of $\Phi_{ii}$ with $n$ and 
$N$ has to be taken into account.

\begin{figure}
\epsfig{file=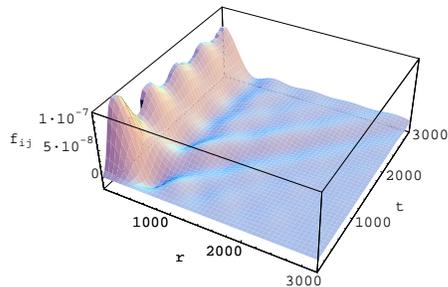,width=5.9cm,angle=0}
\caption{(Color online) The function $\hat{f}(t,r_{ij},\vartheta)$, for
  $\kappa=0.01$ and $\vartheta=\pi/2$ at $T=0$. The limit $r\to 0$ represents
  the single--qubit decoherence, $f_{ii}(t)$. \\[0.2cm]} 
\label{fig.fij}       
\end{figure}

\begin{figure}
\epsfig{file=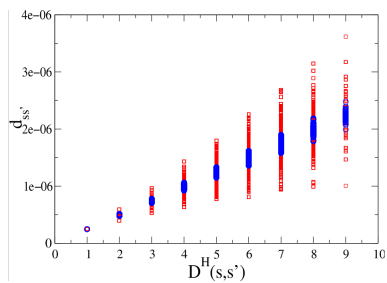,width=5cm,angle=0}
\caption{(Color online) Decoherences $d_{\bs,\bs'}(t)$ of 9 qubits in a
  $3\times 3$ square 
  optical lattice at $t=200.0$ as function of the Hamming distances $D^H(\bs,\bs')$.
Small red squares are for $a=580d$, large blue circles for $a=1000d$. Same
  parameters as in FIG.\ref{fig.fij}.\\[0.2cm] }\label{fig.dofd}       
\end{figure}

The DMT can be evaluated explicitly for
the present physical model. We will
limit ourselves to direct decoherence of  
$n=N$ two level atoms interacting with BBR, such that $\Phi_{ij}=0$,
delegating indirect decoherence to 
another publication.
We express all lengths in terms
of the dipole length $d$ and times in units 
of $d/c$, $t\to ct/d$, and define
$r_{ij}=|\bR_i-\bR_j|/d$. To simplify the
presentation we will 
assume that all dipoles are oriented in the same direction. This means that
$g_k^{(i)}$ is independent of $i$, and $f_{ij}(t,\bR_i-\bR_j)$ 
depends on $i$ and $j$ in the frame aligned with $\bR_i-\bR_j$
only through
$r_{ij}$ and the angle $\vartheta$ between
the $\hat{u}_k^{(i)}$ and 
$\bR_i-\bR_j$. The azimuthal angle is irrelevant, and we can write
$f_{ij}(t,\bR_i-\bR_j)=\hat{f}(t,r_{ij},\vartheta)$.  
In the continuum limit $V\to\infty$, the sums in eq.(\ref{phij})
become integrals over $\bk$ which can be done analytically for $T=0$. The
result reads
$\hat{f}(t,r,\vartheta)=\frac{\alpha}{2\pi}\frac{1}{r^2}\left(   
s(t,r)\sin^2\vartheta+c(t,r)\cos^2\vartheta\right)$ with
\begin{widetext}
\begin{eqnarray}
s(t,r)&=&\frac{2}{1-(t/r)^2}\Big(\cos(\kappa r)\cos(\kappa
t)+\frac{t}{r}\sin(\kappa r)\sin(\kappa
t)-\left(\frac{t}{r}\right)^2\Big) \nonumber\\ 
&&+2\left(\frac{\sin\kappa r}{\kappa r}(1-\cos(\kappa t))-\cos(\kappa
r)\right)+\frac{t}{r}\left(\ln\left|\frac{t-r}{t+r}\right|-{\rm
  Ci}(\kappa|t-r|) +{\rm
  Ci}(\kappa|t+r|)\right)\label{s}\\
c(t,r)&=&2\frac{t}{r}\left({\rm Ci}(\kappa|t-r|)-{\rm
  Ci}(\kappa|t+r|)-\ln\left|\frac{t-r}{t+r}\right|\right)
+\frac{4}{\kappa r}\sin(\kappa r)(\cos(\kappa t)-1)\,,\label{c}
\end{eqnarray} 
\end{widetext}
where $\alpha=e^2/(4\pi\varepsilon_0\hbar c)\simeq 1/137.06$ is the
fine-structure constant, and 
$\kappa=k_{\rm max} d$ a UV cut--off of the integral over $\bk$. A cut--off is
only needed for the diagonal part $f_{ii}$ which corresponds to $r_{ij}=0$,
but in order to assure the  
positivity of the decoherence metric, the same cut--off should be used for
both diagonal and off-diagonal parts. 
The off-diagonal terms $f_{ij}$
arise due to interference effects between two qubits. They decay as
functions of $r$ on a length scale of 
the order $1/\kappa$ (see FIG.\ref{fig.fij}). There can also be regions
in the 
$t,r$ plane where 
$f_{ij}$ becomes negative (not withstanding the fact that the
decoherence metric
is always non--negative).  
One such zone is a ridge along the
light-cone $t=r$. Another one exists for fixed $r$ almost
independently of $t$, for distances of the order $1/\kappa$.  

The function
$\hat{f}(t,r,\vartheta)$ is continuous in the limit 
$r\to 0$, and leads to diagonal matrix elements $f_{ii}$ 
independent of $\vartheta$ and $i$, $f_{ii}(t,{\bf
  0})=\hat{f}(t,0,\vartheta)\equiv f_{ii}(t)$, which describe single--qubit
decoherence (i.e.~$n=N=1$). Explicitly, 
\begin{equation} \label{fii}
f_{ii}(t)=\frac{2}{3\pi}\alpha\left(\frac{\kappa^2}{2}+\frac{1-\cos(\kappa
t)-\kappa t\sin(\kappa t)}{t^2}\right) \,.
\end{equation}
The expressions (\ref{s},\ref{c},\ref{fii}) are exact in the limit $T\to
0$. Corrections due to 
finite temperature are of order $k_BT/(\hbar\omega_{\rm max})$ with
$\omega_{\rm max}=c k_{\rm max}$, and will be
neglected in the following. They can be easily taken into account by
evaluating the integrals obtained from eqs.(\ref{phij}) numerically.
The divergence of $f_{ii}$ for $\kappa\to\infty$ makes the cut--off a
relevant physical quantity. A cut--off always arises, at the latest at
$\kappa\sim 1$ when the dipole approximation breaks down. Much smaller
values of $\kappa$ are expected 
for realistic atoms when transitions to 
excited levels come into play.

It is clear from eq.(\ref{dd}) that in the limit of
large qubit distances ($r\gg 1/\kappa$) the 
decoherence between two code words is just given by the Hamming distance
times the single--qubit decoherence,
$d_{\bs\bs'}(t)=4f_{00}(t)D^H(\bs,\bs')$, as the qubits are then independent
and each qubit contributes to 
decoherence if and only if it is in a superposition of $|-1\rangle$ and
$|1\rangle$. This is shown in  FIG.\ref{fig.dofd} for the example of
$3\times 3$ qubits arranged on a 2D square optical lattice with lattice
constant $a$. The larger $a$ the smaller the spread of the decoherences
around the value given by the Hamming distance and single qubit decoherence.

\begin{figure}
\epsfig{file=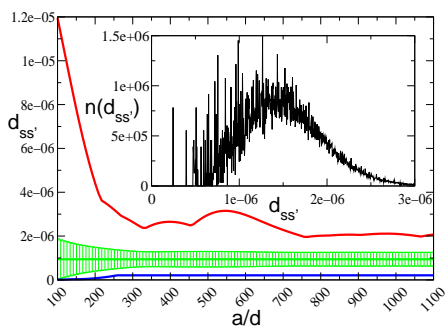,width=5cm,angle=270}
\caption{(Color online) Statistical analysis of all $2^{23}$ independent
  decoherences of $3\times 4$ 
  qubits in a 2D square lattice for $a=100d\ldots 1100d$ (parameters as 
  in FIG.\ref{fig.dofd}). The average decoherence,
  given by the trace of the DMT, is independent of the
  lattice spacings (central green curve, error bars represent $\pm$ one
  standard deviation). 
  The minimum decoherence (bottom curve, blue) indicates a DFS at small
  spacings; the top curve (red) is the maximum decoherence. The inset shows
  a histogram of all decoherences for $a=580d$.  
  }\label{fig.stat}        
\end{figure}

The minimum and maximum decoherence, and the
width of the decoherence distribution, show a substantial dependence on
the lattice spacing (see FIG.\ref{fig.stat}). 
Notably, for small lattice
spacing ($r\ll 1/\kappa$) the minimum decoherence tends to zero. A DFS
arises, as 
the atoms then couple in a symmetric way 
to the heat bath (the latter does not contain sufficiently short waves for
distinguishing the positions of the atoms). The 
decoherence metric allows to generalize 
the DFS concept in the sense that even for non--symmetric coupling one may
optimize now the performance of the quantum computer by encoding in the
states with {\em smallest} decoherence, which can be explicitly found using the
decoherence metric. In \cite{Beige00} a method was introduced that allows to
quantum compute within a DFS, based on a strong coupling to the environment
and corresponding broadening of non--DFS states. It is expected that this
method can be generalized to the current approach.

As a summary,
I have derived a ``decoherence metric'' that induces a natural distance
between binary quantum code words. The decoherence metric
generalizes the well-known Hamming 
distance and determines directly the time--dependent decoherence in an
arbitrarily large system of qubits with degenerate energy levels coupled to
a heat bath of harmonic osciallators. \\
{\em Acknowledgments:}
I wish to thank Olivier Giraud for 
valuable discussions. This work
was supported by the Agence 
National de la Recherche 
(ANR), project INFOSYSQQ.
\bibliography{../mybibs_bt}

\end{document}